\documentclass[onecolumn,showpacs]{revtex4}

\usepackage{graphicx}
\usepackage{dcolumn}
\usepackage{bm}

\input epsf

\begin{document}

\title{\Large Role of Tachyonic Field in Accelerating Universe in Presence of Perfect Fluid}

\author{\bf Writambhara Chakraborty$^1$\footnote{writam1@yahoo.co.in}
and Ujjal Debnath$^2$\footnote{ujjaldebnath@yahoo.com} }

\affiliation{$^1$Department of Mathematics, New Alipore College, New Alipore L Block, Kolkata-700 053, India.\\
$^2$Department of Mathematics, Bengal Engineering and Science
University, Shibpur, Howrah-711 103, India. }

\date{\today}

\begin{abstract}
Recently tachyonic field has been depicted as dark energy model
to represent the present acceleration of the Universe. In this
paper we have considered mixture of tachyonic fluid with a
perfect fluid. For this purpose we have considered barotropic
fluid and Generalized Chaplygin gas(G.C.G.). We have considered a
particular form of the scale factor. We have solved the equations
of motion to get the exact solutions of the density, tachyonic
potential and the tachyonic field. We have introduced a coupling
term to show that the interaction decays with time. Also we have
shown that the nature of the potentials vary so as the
interaction term reduces the potential in both the cases.
\end{abstract}


\maketitle

\section{\normalsize\bf{Introduction}}

The observational confirmation about the accelerated expansion of
the Universe [1, 2] has given rise to a lot of dark energy models
[3-6], which are supposed to be the reason behind this present
acceleration. This mysterious fluid called dark energy is
believed to dominate over the matter content of the Universe by
70 $\%$ and to have enough negative pressure as to drive present
day acceleration. Most of the dark energy models involve one or
more scalar fields with various actions and with or without a
scalar field potential [7]. Recently tachyonic field with
Lagrangian ${\cal {L}}=-V(T)\sqrt{1+g^{\mu \nu}~\partial{_{\mu}}T
\partial{_{\nu}} T}$ [8] has gained a lot of importance in this
respect. The energy-momentum tensor of the tachyon field can be
seen as a combination of two fluids, dust with pressure zero and
a cosmological constant with $p=-\rho$, thus generating enough
negative pressure such as to drive acceleration. Also the tachyon
field has a potential which has an unstable maximum at the origin
and decays to almost zero as the field goes to infinity.
Depending on various forms of this potential following this
asymptotic behaviour a lot of works have been carried out on
tachyonic dark energy [9, 10], tachyonic dark matter [11, 12] and
inflation models [13].\\

To obtain a suitable evolution of the Universe an interaction is
often assumed such that the decay rate should be proportional to
the present value of the Hubble parameter for good fit to the
expansion history of the Universe as determined by the Supernovae
and CMB data [14]. These kind of models describe an energy flow
between the components so that no components are conserved
separately. An interacting tachyonic-dark matter model has been
studied in ref. [15].\\

In this paper, we consider a model which comprises of a two
component mixture. Firstly we consider a mixture of barotropic
fluid with tachyonic field without any interaction between them,
so that both of them retain their properties separately. Then we
consider an energy flow between them by introducing an
interaction term which is proportional to the product of the
Hubble parameter and the density of the barotropic fluid. We show
that the energy flow being considerably high at the beginning
falls down noticeably with the evolution of the Universe
indicating a more stable situation. Also in both the cases we
find the exact solutions for  the tachyonic field and the
tachyonic potential and show that the tachyonic potential follows
the asymptotic behaviour discussed above. Here the tachyonic
field behaves as the dark energy component whereas the dust acts
as the cold dark matter. Next we consider tachyonic dark matter,
the Generalized Chaplygin Gas (GCG) being the dark energy
component. GCG, identified by the equation of state (EOS)
$p=-B/\rho^{\alpha}$ with $0\le \alpha \le 1$ has been considered
as a suitable dark energy model by several authors [16, 17]. Here
we consider the mixture of GCG with tachyonic dark matter. Later
we have also considered an interaction between these two fluids
by introducing a coupling term which is proportional to the
product of Hubble constant and the energy density of the GCG. The
coupling function decays with time indicating a strong energy flow
at the initial period and weak interaction at later stage
implying a stable situation. Like the previous case we have found
the exact solution of the tachyonic potential. To keep the
observational support of recent acceleration we have considered a
particular form of evolution of the Universe here as
\begin{equation}
a=t^{n}
\end{equation}
such that the deceleration parameter reads $q=-\frac{a
\ddot{a}}{{\dot{a}}^{2}}=-(1-\frac{1}{n})$, where $a$ is the scale
factor. Hence for $n>1$ we always get an accelerated expansion
and for $n=1$ we get a constant expansion of the Universe. This
kind of recipe has been studied in ref. [12].\\

The paper is organized as follows:\\

Section II deals with the field equations of the tachyonic field
and Einstein field equations. In sections III and IV we have
considered models represented by mixture of tachyonic field with
barotropic fluid and GCG respectively. These sections are each
subdivided into two parts showing the effect of these models with
or without interaction. We have taken some particular values of
the parameters and constants for the graphical representation of
the tachyonic potential. The paper
ends with a short discussion in section V.\\

\section{\normalsize\bf{Tachyonic Fluid Model and Einstein Field Equations}}

The action for the homogeneous tachyon condensate of string
theory in a gravitational background is given by,
\begin{equation}
S=\int {\sqrt{-g}~ d ^{4} x \left[\frac{\cal R}{16 \pi G}+{\cal
L}\right]}
\end{equation}
where $\cal L$ is the Lagrangian density given by,
\begin{equation}
{\cal {L}}=-V(T)\sqrt{1+g^{\mu \nu}~\partial{_{\mu}}T
\partial{_{\nu}} T}
\end{equation}
where $T$ is the tachyonic field, $V(T)$ is the tachyonic
potential and $\cal R$ is the Ricci Scalar. The energy-momentum
tensor for the tachyonic field is,

\begin{eqnarray}
\begin{array} {ccc}
T_{\mu \nu}=-\frac{2 \delta S}{\sqrt{-g}~ \delta g^{\mu
\nu}}=-V(T)\sqrt{1+g^{\mu \nu} \partial _{\mu}T \partial_{\nu}
T}g^{\mu \nu}+V(T) \frac{\partial _{\mu}T \partial_{\nu}
T}{\sqrt{1+g^{\mu \nu} \partial _{\mu}T \partial_{\nu} T}}\\\\\
=p_{T}~g_{\mu \nu}+(p_{T}+\rho_{T})u_{\mu} u_{\nu}\\\\
\end{array}
\end{eqnarray}

where the velocity $u_{\mu}$ is :
\begin{equation}
u_{\mu}=-\frac{\partial_{\mu}T}{\sqrt{-g^{\mu \nu} \partial
_{\mu}T
\partial_{\nu} T}}
\end{equation}

with $u^{\nu} u_{\nu}=-1$.\\

The energy density $\rho_{T}$ and the pressure $p_{T}$ of the
tachyonic field therefore are,
\begin{equation}
\rho_{T}=\frac{V(T)}{\sqrt{1-{\dot{T}}^{2}}}
\end{equation}
and
\begin{equation}
p_{T}=-V(T) \sqrt{1-{\dot{T}}^{2}}
\end{equation}

Hence the EOS parameter of the tachyonic field becomes,
\begin{equation}
\omega_{T}=\frac{p_{T}}{\rho_{T}}=-(1-{\dot{T}}^{2})
\end{equation}
and
\begin{equation}
p_{T} \rho_{T}=-{V^{2}(T)}
\end{equation}
which represents pure Chaplygin gas if $V(T)$ is constant.\\

Now the metric of a spatially flat isotropic and homogeneous
Universe in FRW model is,

\begin{equation}
ds^{2}=dt^{2}-a^{2}(t)\left[dr^{2}+r^{2}(d\theta^{2}+sin^{2}\theta
d\phi^{2})\right]
\end{equation}

where $a(t)$ is the scale factor.\\

The Einstein field equations are (choosing $8\pi G=c=1$)

\begin{equation}
3\frac{\dot{a}^{2}}{a^{2}}=\rho_{tot}
\end{equation}
and
\begin{equation}
6\frac{\ddot{a}}{a}=-(\rho_{tot}+3p_{tot})
\end{equation}

where, $\rho_{tot}$ and $p_{tot}$ are the total energy density and
the pressure of the Universe.\\

The energy conservation equation is
\begin{equation}
\dot{\rho}_{tot}+3\frac{\dot{a}}{a}(\rho_{tot}+p_{tot})=0
\end{equation}

\section{\normalsize\bf{Mixture of Tachyonic dark energy and
barotropic fluid}}

Now we consider a two fluid model consisting of tachyonic field
and barotropic fluid. The EOS of the barotropic fluid is given by,
\begin{equation}
p_{b}=\omega_{b} \rho_{b}
\end{equation}

where $p_{b}$ and $\rho_{b}$ are the pressure and energy density
of the barotropic fluid. Hence the total energy density and
pressure are respectively given by,
\begin{equation}
\rho_{tot}=\rho_{b}+\rho_{T}
\end {equation}
and
\begin{equation}
p_{tot}=p_{b}+p_{T}
\end {equation}

\subsection{\normalsize\bf{Without Interaction}}

First we consider that the two fluids do not interact with each
other so that they are conserved separately. Therefore, the
conservation equation (13) reduces to,
\begin{equation}
\dot{\rho}_{T}+3\frac{\dot{a}}{a}(\rho_{T}+p_{T})=0
\end{equation}
and
\begin{equation}
\dot{\rho}_{b}+3\frac{\dot{a}}{a}(\rho_{b}+p_{b})=0
\end{equation}

Equation (18) together with equation (14) give,
\begin{equation}
\rho_{b}=\rho_{0}~a^{-3(1+\omega_{b})}
\end{equation}
Now, we consider a power law expansion of the scale factor $a(t)$
given by equation (1).\\
\\
Using (1), equation (19) reduces to,
\begin{equation}
\rho_{b}=\rho_{0}~t^{-3n(1+\omega_{b})}
\end{equation}
Also the energy density corresponding to the tachyonic field
becomes,
\begin{equation}
\rho_{T}=\frac{1}{t^{2}}
\left[3n^{2}-\rho_{0}~t^{-3n(1+\omega_{b})+2}\right]
\end{equation}
Solving the equations the tachyonic field is obtained as,
\begin{equation}
T=\sqrt{1+\omega_{b}}t ~Appell~ F_{1}
\left[\frac{1}{3(1+\omega_{b})n-2}, \frac{1}{2}, -\frac{1}{2},
1+\frac{1}{3(1+\omega_{b})n-2}, \frac{3n^{2}}{\rho_{0}}
t^{3(1+\omega_{b})n-2}, \frac{2n}{\rho_{0} (1+\omega_{b})}
t^{3(1+\omega_{b})n-2} \right]
\end{equation}
where, $Appell ~F1[a,b_{1},b_{2},c,x,y]$ is the Appell
Hypergeometric function of two variables $x$ and $y$.\\

Also the potential will be of the form,
\begin{equation}
V(T)= \sqrt{\frac{3n^{2}}{t^{2}}-\rho_{0}~t^{-3n(1+\omega_{b})}}
\sqrt{\frac{3n^{2}}{t^{2}}-\frac{2n}{t^{2}}+\omega_{b}\rho_{0}~t^{-3n(1+\omega_{b})}}
\end{equation}

We can show the graphical representation of the potential against
time in figure 1. We can see that $V \rightarrow 0$ with time,
thus retaining the original property of the tachyon potential.\\
\begin{figure}
\includegraphics[height=2.2in]{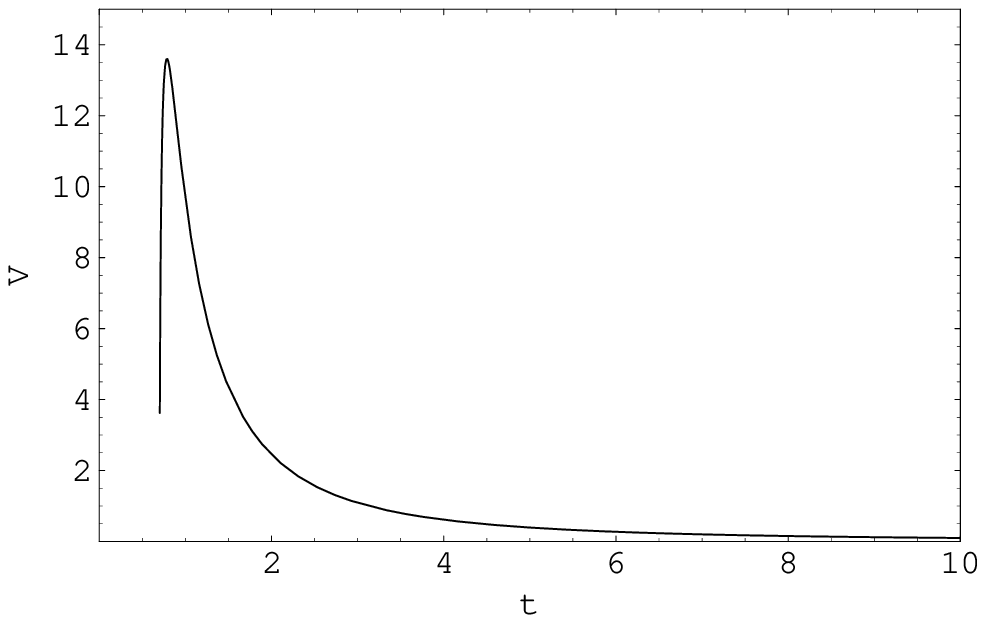}

Fig.1 \vspace{3mm}

\vspace{5mm} Fig. 1 shows the variation of $V$  against $t$ for $
n=2, \rho_{0}=1, \omega_{b}=\frac{1}{2}$. \hspace{14cm}
\vspace{4mm}

\end{figure}

\subsection{\normalsize\bf{With Interaction}}

Now we consider an interaction between the tachyonic field and
the barotropic fluid by introducing a phenomenological coupling
function which is a product of the Hubble parameter and the
energy density of the barotropic fluid. Thus there is an energy
flow between the two fluids. \\

Now the equations of motion corresponding to the tachyonic field
and the barotropic fluid are respectively,
\begin{equation}
\dot{\rho}_{T}+3\frac{\dot{a}}{a}(\rho_{T}+p_{T})=-3H\delta
\rho_{b}
\end{equation}
and
\begin{equation}
\dot{\rho_{b}}+3\frac{\dot{a}}{a}(\rho_{b}+p_{b})=3H\delta
\rho_{b}
\end{equation}
 where $\delta$ is a coupling constant.\\

Solving equation (25) with the help of equation (14), we get,
\begin{equation}
\rho_{b}=\rho_{0}~a^{-3(1+\omega_{b}-\delta)}
\end{equation}
Considering the power law expansion (1), we get
\begin{equation}
\rho_{b}=\rho_{0}~t^{-3n(1+\omega_{b})-\delta}
\end{equation}
Equation (11) and (27) give,
\begin{equation}
\rho_{T}=\frac{3n^{2}}{t^{2}}
-\rho_{0}~t^{-3n(1+\omega_{b})-\delta}
\end{equation}
Solving the equations the tachyonic field is obtained as,
\begin{equation}
T=\sqrt{1+\omega_{b}}t ~Appell~ F_{1}
\left[\frac{1}{3(1+\omega_{b}-\delta)n-2}, \frac{1}{2},
-\frac{1}{2}, 1+\frac{1}{3(1+\omega_{b}-\delta)n-2},
\frac{3n^{2}}{\rho_{0}} t^{3(1+\omega_{b}-\delta)n-2},
\frac{2n}{\rho_{0} (1+\omega_{b})} t^{3(1+\omega_{b}-\delta)n-2}
\right]
\end{equation}
Also the potential will be of the form,
\begin{equation}
V(T)=
\sqrt{\frac{3n^{2}}{t^{2}}-\rho_{0}~t^{-3n(1+\omega_{b}-\delta)}}
\sqrt{\frac{3n^{2}}{t^{2}}-\frac{2n}{t^{2}}+\omega_{b}\rho_{0}~t^{-3n(1+\omega_{b}-\delta)}}
\end{equation}

In this case also $V\rightarrow 0$ with time as shown in the
graphical representation of $V$ in figure 2.\\

\begin{figure}
\includegraphics[height=2.2in]{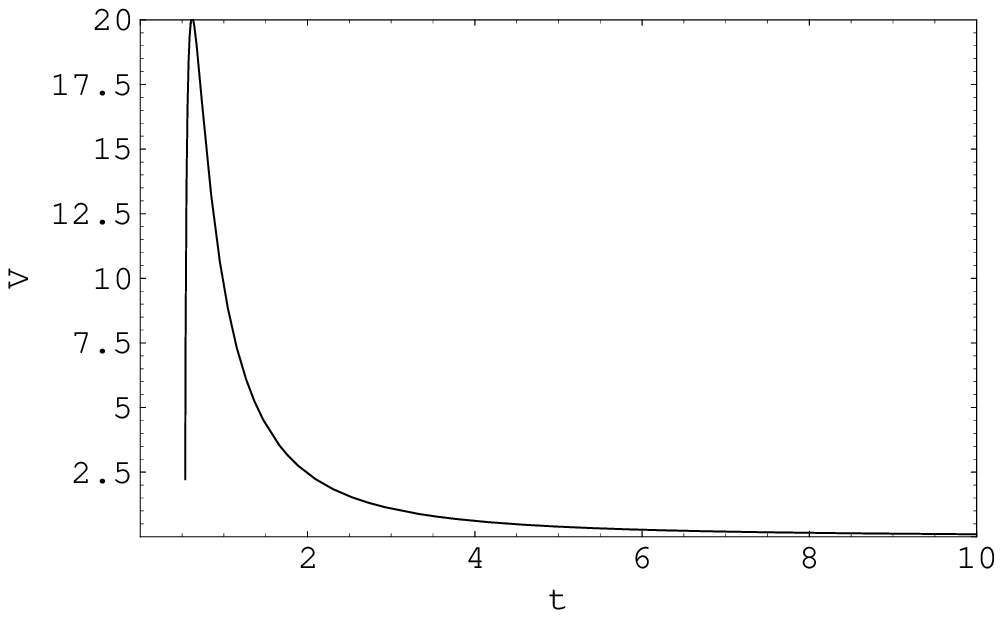}

Fig.2 \vspace{3mm}

\vspace{5mm} Fig. 2 shows the variation of $V$  against $t$ for $
n=2, \rho_{0}=1, \omega_{b}=\frac{1}{2}, \delta=\frac{1}{2}$.
\hspace{14cm} \vspace{4mm}

\end{figure}

\section{\normalsize\bf{Mixture of Tachyonic dark matter and Generalized Chaplygin gas}}

Now we consider a two fluid model consisting of tachyonic field
and G.C.G. The EOS of G.C.G. is given by,
\begin{equation}
p_{ch}=-B/{\rho}_{ch}^{\alpha}~~~~~~~~~~~\text~~~~~~~~~~~~ 0\le
\alpha \le 1, B>0.
\end{equation}

where $p_{ch}$ and $\rho_{ch}$ are the pressure and energy density
of G.C.G. Hence the total energy density and pressure are
respectively given by,
\begin{equation}
\rho_{tot}=\rho_{ch}+\rho_{T}
\end {equation}
and
\begin{equation}
p_{tot}=p_{ch}+p_{T}
\end {equation}

\subsection{\normalsize\bf{Without Interaction}}

First we consider that the two fluids do not interact with each
other so that they are conserved separately. Therefore, the
conservation equation (13) reduces to,
\begin{equation}
\dot{\rho}_{T}+3\frac{\dot{a}}{a}(\rho_{T}+p_{T})=0
\end{equation}
and
\begin{equation}
\dot{\rho}_{ch}+3\frac{\dot{a}}{a}(\rho_{ch}+p_{ch})=0
\end{equation}

Equation (35) together with equation (31) give,
\begin{equation}
\rho_{ch}=\left[B+\frac{\rho_{00}}{a^{3(1+\alpha)}}\right ]^{\frac{1}{(1+\alpha)}}
\end{equation}

Using (1), equation(36) reduces to,
\begin{equation}
\rho_{ch}=[B+\rho_{00}t^{-3n(1+\alpha)}]^{\frac{1}{(1+\alpha)}}
\end{equation}
Hence the energy density of the tachyonic fluid is,
\begin{equation}
\rho_{T}=\frac{3n^{2}}{t^{2}}
-[B+\rho_{00}t^{-3n(1+\alpha)}]^{\frac{1}{(1+\alpha)}}
\end{equation}
Solving the equations the tachyonic field and the tachyonic
potential are obtained as,
\begin{equation}
T=\int\sqrt{\frac{\frac{2n}{t^{2}}-\rho_{00} t^{-3n(1+\alpha)}
[B+\rho_{00}t^{-3n(1+\alpha)}]^{-\frac{\alpha}{(1+\alpha)}}}{\frac{3n^{2}}{t^{2}}
-[B+\rho_{00}t^{-3n(1+\alpha)}]^{\frac{1}{(1+\alpha)}}}}dt
\end{equation}
Also the potential will be of the form,
\begin{equation}
V(T)=
\sqrt{\frac{3n^{2}}{t^{2}}-[B+\rho_{00}t^{-3n(1+\alpha)}]^{\frac{1}{(1+\alpha)}}}
\sqrt{\frac{3n^{2}}{t^{2}}-\frac{2n}{t^{2}}-B[B+\rho_{00}t^{-3n(1+\alpha)}]^{-\frac{\alpha}{(1+\alpha)}}}
\end{equation}

\begin{figure}
\includegraphics[height=2.2in]{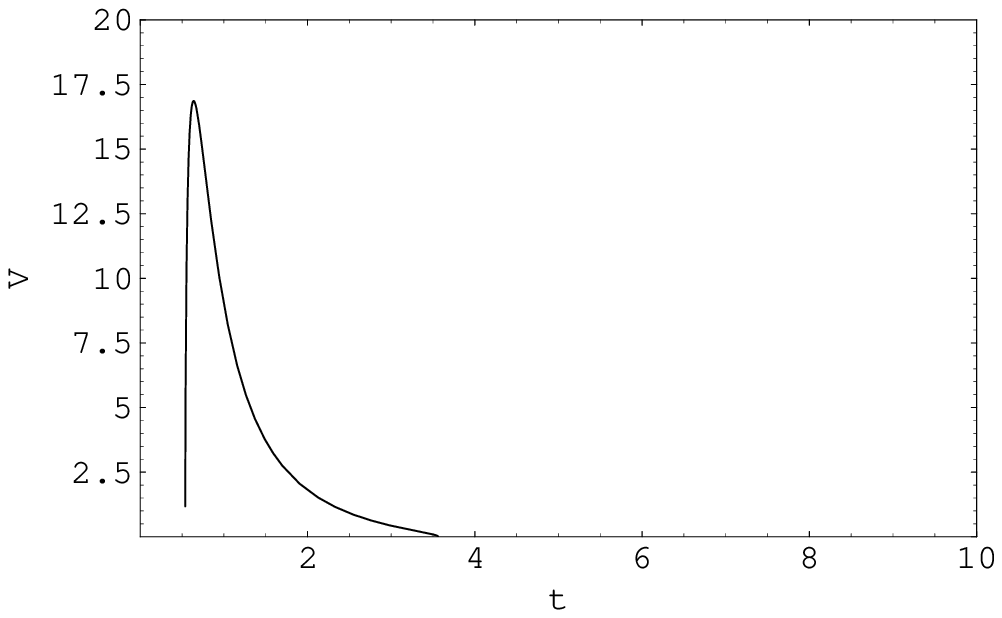}

Fig.3 \vspace{3mm}

\vspace{5mm} Fig. 3 shows the variation of $V$  against $t$ for $
B=\frac{1}{2}, n=2, \rho_{00}=1, \alpha=\frac{1}{2}$.
\hspace{14cm} \vspace{4mm}

\end{figure}

Like the mixture of tachyonic fluid with barotropic fluid in this
case also the potential $V$ starting from a low value increases
largely and then decreases to $0$ with time as shown in figure 3.\\

\subsection{\normalsize\bf{With Interaction}}

Now we consider an interaction between the tachyonic fluid and
G.C.G. by phenomenologically introducing an interaction term as a
product of the Hubble parameter and the energy density of the
Chaplygin gas. Thus there is an energy
flow between the two fluids. \\

\begin{figure}
\includegraphics[height=2.2in]{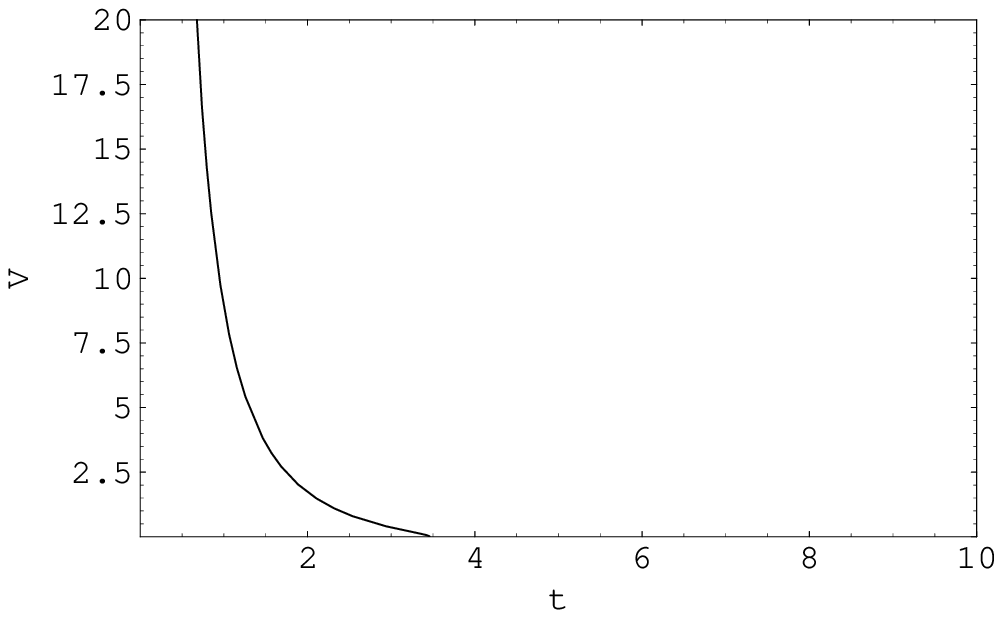}

Fig.4 \vspace{3mm}

\vspace{5mm} Fig. 4 shows the variation of $V$  against $t$ for $
B=\frac{1}{2}, n=2, \rho_{00}=1, \alpha=\frac{1}{2},
\epsilon=\frac{1}{2}$. \hspace{14cm} \vspace{4mm}

\end{figure}

Now the equations of motion corresponding to the tachyonic field
and G.C.G. are respectively,
\begin{equation}
\dot{\rho}_{T}+3\frac{\dot{a}}{a}(\rho_{T}+p_{T})=-3H\epsilon
\rho_{ch}
\end{equation}
and
\begin{equation}
\dot{\rho}_{ch}+3\frac{\dot{a}}{a}(\rho_{ch}+p_{ch})=3H\epsilon
\rho_{ch}
\end{equation}
where $\epsilon$ is a coupling constant.\\

Solving equation (42) with the help of equation (31) and (1), we
get,
\begin{equation}
\rho_{ch}=\left[\frac{B}{1-\epsilon}+\rho_{00}
t^{-3n(1+\alpha)(1-\epsilon)}\right]^{\frac{1}{(1+\alpha)}}
\end{equation}

Also the energy density of the tachyonic field will be read as,
\begin{equation}
\rho_{T}=\frac{3n^{2}}{t^{2}}
-\left[\frac{B}{1-\epsilon}+\rho_{00}
t^{-3n(1+\alpha)(1-\epsilon)}\right]^{\frac{1}{(1+\alpha)}}
\end{equation}
Solving the equations the tachyonic field is obtained as,
\begin{equation}
T=\int\sqrt{\frac{\frac{2n}{t^{2}}-\left[\frac{B \epsilon
}{1-\epsilon}+\rho_{00} t^{-3n(1+\alpha)(1-\epsilon)}\right]
\left[\frac{B}{1-\epsilon}+\rho_{00}
t^{-3n(1+\alpha)(1-\epsilon)}\right]^{-\frac{\alpha}{(1+\alpha)}}}{\frac{3n^{2}}{t^{2}}
-\left[\frac{B}{1-\epsilon}+\rho_{00}
t^{-3n(1+\alpha)(1-\epsilon)}\right]^{\frac{1}{(1+\alpha)}}}}dt
\end{equation}
Also the potential will be of the form,
\begin{equation}
V(T)=\sqrt{\frac{3n^{2}}{t^{2}}-\left[\frac{B}{1-\epsilon}+\rho_{00}
t^{-3n(1+\alpha)(1-\epsilon)}\right]^{\frac{1}{(1+\alpha)}}}
\sqrt{\frac{3n^{2}}{t^{2}}-\frac{2n}{t^{2}}-B\left[\frac{B}{1-\epsilon}+\rho_{00}
t^{-3n(1+\alpha)(1-\epsilon)}\right]^{-\frac{\alpha}{(1+\alpha)}}}
\end{equation}

In this case the potential starting from a large value tends to
$0$ (figure 4).\\

\section{\normalsize\bf{Discussion}}

We have considered the flat FRW Universe driven by a mixture of
tachyonic field and a perfect fluid. We have considered
barotropic fluid and Chaplygin gas for this purpose. We have
presented accelerating expansion of our Universe due to
interaction/without interaction of the mixture of these fluids.
We have found the exact solution of the density and potential by
considering a power law expansion of the scale factor. We show
that these potentials represent the same decaying nature
regardless the interaction between the concerned fluids. Since we
have considered a power law expansion of the scale factor of the
form $a=t^{n}$, we see that for the present acceleration of the
Universe to support the observational data we need $n>1$. Now we
consider the interaction terms between these fluids. For the
mixture of barotropic fluid with tachyonic fluid, we see that the
interaction term reduces the potential. Also for the mixture of
G.C.G. with tachyonic fluid the interaction parameter $\epsilon$
satisfying  $0<\epsilon<1$ so that equation (43) exists for
smaller values of $t$. In this case also the interaction reduces
the potential. Also if we consider only tachyonic fluid with the
power law expansion, we see that the potential (which is obtained
to be $V=\frac{3 n^{2}}{t^{2}} \sqrt{1-\frac{2}{3n}}$) is greater
than that we get in mixtures. Also the potentials differ in the
two cases we have considered. For the mixture with G.C.G. the
potential decreases faster than that in case of mixture with
barotropic fluid.\\\\

{\bf Acknowledgement:}\\

The authors are thankful to IUCAA, India for warm hospitality
where part of the work was carried out. Also UD is thankful to
UGC, Govt. of India for providing research project grant (No. 32-157/2006(SR)).\\

{\bf References:}\\
\\
$[1]$  N. A. Bachall, J. P. Ostriker, S. Perlmutter and P. J.
Steinhardt, {\it Science} {\bf 284} 1481 (1999).\\
$[2]$ S. J. Perlmutter et al, {\it Astrophys. J.} {\bf 517} 565
(1999).\\
$[3]$ V. Sahni and A. A. Starobinsky, {\it Int. J. Mod. Phys. A}
{\bf 9} 373 (2000).\\
$[4]$ P. J. E. Peebles and B. Ratra, {\it Rev. Mod. Phys.} {\bf
75} 559 (2003).\\
$[5]$ T. Padmanabhan, {\it Phys. Rept.} {\bf 380} 235 (2003).\\
$[6]$ E. J. Copeland, M. Sami, S. Tsujikawa, {\it Int. J. Mod.
Phys. D} {\bf  15} 1753 (2006).\\
$[7]$ I. Maor and R. Brustein, {\it Phys. Rev. D} {\bf 67} 103508
(2003); V. H. Cardenas and S. D. Campo, {\it Phys. Rev. D} {\bf
69} 083508 (2004); P.G. Ferreira and M. Joyce, {\it Phys.
Rev. D} {\bf 58} 023503 (1998).\\
$[8]$ A. Sen, {\it JHEP} {\bf04} 048 (2002); A. Sen, {\it JHEP}
{\bf 07} 065 (2002); A. Sen, {\it Mod. Phys. Lett. A} {\bf 17}
1797(2002).\\
$[9]$ J. S. Bagla, H. K. Jassal and T. Padmanabhan, {\it Phys.
Rev. D} {\bf 67} 063504 (2003); E. J. Copeland, M. R. Garousi, M.
Sami and S. Tsujikawa, {\it Phys. Rev D} {\bf 71} 043003 (2005);
G. Calcagni and A. R. Liddle, {\it astro-ph}/0606003.\\
$[10]$ E. J. Copeland, M. Sami and S. Tsujikawa, {\it Int. J.
Mod. Phys. D} {\bf 15} 1753 (2006).\\
$[11]$ A. DeBenedictis, A. Das and S. Kloster, {\it Gen. Rel.
Grav.} {\bf 36} 2481 (2004); A. Das, S. Gupta, T. D. Saini and S.
Kar, {\it Phys. Rev D} {\bf 72} 043528 (2005).\\
$[12]$ T. Padmanabhan, {\it Phys. Rev. D} {\bf 66} 021301 {\bf R}
(2002).\\
$[13]$ M. Sami, {\it Mod. Phys. Lett. A} {\bf 18} 691 (2003); B.
C. Paul and D. Paul, {\it Int. J. Mod. Phys. D} {\bf 14} 1831
(2005).\\
$[14]$ M. S. Berger, H. Shojaei, {\it Phys. rev. D} {\bf 74}
043530 (2006).\\
$[15]$ R. Herrera, D. Pavon, W. Zimdahl, {\it gen. Rel. Grav.}
{\bf 36} 2161 (2004).\\
$[16]$ V. Gorini, A. Kamenshchik and U. Moschella, {\it Phys. Rev.
D} {\bf 67} 063509 (2003); U. Alam, V. Sahni , T. D. Saini and
A.A. Starobinsky, {\it Mon. Not. Roy. Astron. Soc.} {\bf 344}, 1057 (2003).\\
$[17]$ M. C. Bento, O. Bertolami and A. A. Sen, {\it Phys. Rev. D}
{66} 043507 (2002).\\

\end{document}